\documentstyle[pre,preprint,aps]{revtex}
\tightenlines
\newcommand{\siml}{\stackrel{<}{\sim}}
\newcommand{\simg}{\stackrel{>}{\sim}}

\begin{document}
\draft

\title{
An augmented moment method for 
stochastic ensembles \\
with delayed couplings: 
I. Langevin model 
}
\author{
Hideo Hasegawa
\footnote{e-mail:  hasegawa@u-gakugei.ac.jp}
}
\address{
Department of Physics, Tokyo Gakugei University,
Koganei, Tokyo 184-8501, Japan
}
\date{\today}
\maketitle
\begin{abstract}
By employing a semi-analytical dynamical mean-field 
approximation theory previously proposed by the author
[H. Hasegawa, Phys. Rev. E {\bf 67}, 041903 (2003)],
we have developed an augmented moment method (AMM) in order to
discuss dynamics of an $N$-unit ensemble
described by Langevin equations with delays.
In AMM, original 
$N$-dimensional {\it stochastic} delay differential equations 
(SDDEs) are transformed to infinite-dimensional {\it deterministic} DEs
for means and correlations of local as well as global variables.
Infinite-order DEs arising from the non-Markovian 
property of SDDE, are terminated at the finite level $m$
in the level-$m$ AMM (AMM$m$), which yields
$(3+m)$-dimensional deterministic DEs.
Model calculations have been made 
for linear and nonlinear Langevin models.
The stationary solution of AMM for 
the linear Langevin model with $N=1$
is nicely compared to the exact result.
In the nonlinear Langevin ensemble, 
the synchronization 
is shown to be enhanced near the transition point
between the oscillating and non-oscillating states.
Results calculated by AMM6 
are in good agreement with those obtained 
by direct simulations.

\end{abstract}

\noindent
\vspace{0.5cm}
%
\section{INTRODUCTION}


The time delay plays an important role in many systems
such as optical devices \cite{Ikeda80},
physiological \cite{Mackey79} 
and biological systems \cite{Foss96}.
The effect of time delays has been theoretically studied
by using the time-delay differential equations (DEs). 
Its exposed behavior includes the multistability 
and the bifurcation leading to chaos. 
It is well known that noise also plays important roles
in these systems, and
its effects have been thoroughly investigated
with the use of stochastic DEs.
One of its representative phenomena is 
the stochastic resonance \cite{Gam98},
in which the signal-to-noise ratio
is enhanced for sub-threshold signals.

In real systems, both noises and time delays coexist,
and the combined effect may be described by
stochastic delay differential equation (SDDE).
For instance, SDDEs are used in optics \cite{Gar96}
and physiology \cite{Longtin90} to model noise-driven systems 
exhibiting delay feedback.
In recent years, there has been a growing interest 
in combined effects of noise and delay.
The theory for SDDE remains much less studied and 
has been a subject of several recent papers 
\cite{Kuchler92}-\cite{Tsi01},
in which the stability condition for the equilibrium solution
of linear delay Langevin equation has been studied.
Its stationary solution is investigated
by using the step by step method \cite{Kuchler92}
and the moment method \cite{Mackey95}.
The Fokker-Planck equation (FPE) method is applied to SDDE in the
limit of a small delay \cite{Guil99}. 
These studies have been confined to the stationary solution
of SDDE. More interesting is, however, expected to be 
its dynamics in a stochastic system with a large time delay.

Real physiological and biological 
systems usually consist of many elements, each of
which is described by SDDE.
A typical example is a living brain, in which
a small cluster contains thousands of similar neurons.
Each neuron which is subject to various kinds of noises,
receives spikes from hundreds of other neurons
through dendrites with a transmission delay and generates spikes
propagating along axons.
Theoretical study on such coupled, stochastic systems has been
made by using direct simulations (DSs)
\cite{Kim97,Borisyuk02,Huber03}
and analytical methods like FPE \cite{Zorzano03}. 
Since the time to simulate such systems by conventional methods 
grows as $N^2$ with $N$, the size of the ensemble,
it is rather difficult to simulate systems with the realistic size
of $N \sim 100-1000$.
Although FPE is a powerful method in dealing with the stochastic DE, 
a simple application of FPE to SDDE fails because of 
the non-Markovian property of SDDE: 
an evaluation of the probability density at the time $t$ requires
prior knowledge of the conditional probability density 
between times of $t$ and $t-\tau$, $\tau$ being the delay time.  

Quite recently the present author 
\cite{Hasegawa03}\cite{Hasegawa03b} has proposed 
a dynamical mean-field approximation (DMA)
as a semi-analytical method dealing with
large-scale ensembles subject to noises,
extending the moment method \cite{Rod96}-\cite{Tanabe01}. 
DMA has been first applied to an $N$-unit ensemble described by
the FitzHugh-Nagumo (FN) neuron model without time delays 
\cite{FitzHugh61}, for which original $2N$-dimensional 
stochastic DEs are transformed to eight-dimensional
deterministic DEs for moments of local and global 
variables \cite{Hasegawa03}. In a subsequent paper \cite{Hasegawa03b},
DMA is applied to an $N$-unit general neuron ensemble, each of which
is described by coupled $K$-dimensional DEs,
transforming $KN$-dimensional DEs to $N_{eq}$-dimensional DEs
where $N_{eq}=K(K+2)$.
In the case of the Hodgkin-Huxley (HH) model 
with $K=4$ \cite{Hodgkin52}, 
we get $N_{eq}=24$.
The spiking-time precision and
the synchronization in FN and HH neuron ensembles
have been studied as functions of the noise intensity,
the coupling strength and the ensemble size.
The feasibility of DMA has been demonstrated
in Refs. \cite{Hasegawa03} and \cite{Hasegawa03b}. 

The purpose of the present paper is to apply
DMA to Langevin ensembles with delays, 
which are expected to be good models
representing not only interconnected neural networks
but also social and technological ones. 
When DMA is applied to ensembles described 
by linear and nonlinear Langevin equations with delays,
the original $N$-dimensional stochastic DEs are
transformed to the infinite-dimensional deterministic DEs for 
means and correlation 
functions of local and global variables.
Infinite-order recursive DEs arising from
the non-Markovian property of SDDE, are terminated
at the finite level $m$ in our approximate method,
which is hereafter referred to as
the {\it augmented moment method} (AMM).
We may study dynamics and synchronization
of linear and nonlinear
Langevin ensembles with delayed couplings,
and examine the validity of AMM whose results
are compared to results of DSs.
In particular, for the linear Langevin model with $N=1$,
a comparison is possible with the exact stationary solution
\cite{Kuchler92}.

The paper is organized as follows.
In the next Sec. II, we describe the adopted model and
method to derive
the infinite-dimensional deterministic DEs
from the original $N$-dimensional stochastic DEs.
Infinite-order recursive DEs are terminated
at the finite level $m$ in the level-$m$ AMM
(AMM$m$).
Model numerical calculations for the
linear Langevin model are reported in Sec. IIIA, where calculated
results of AMM6 for $N=1$ are nicely compared to 
exact solutions available for the stationary state \cite{Kuchler92}. 
Our AMM is compared also to the small-delay approximation (SDA)
\cite{Guil99} which is valid for a very small delay.
In Sec. IIIB, we present model calculations for
the nonlinear Langevin model in which the stable oscillation
is induced by an applied spike for an appreciable delay.
The synchronization in the ensemble 
is investigated.
It is shown that results of AMM6 are
in good agreement with those of DSs.
The final Sec. IV is devoted to conclusions and discussions.
In a following paper \cite{Hasegawa04b},
our AMM has been applied to ensembles described 
by the noisy FN neuron model with delayed couplings.

\section{Ensembles described by Langevin model}

\subsection{Basic formulation}

Dynamics of a Lanvegin ensemble
with delayed couplings is assumed to be described by
\begin{equation}
\frac{d x_i(t)}{dt}= F(x_i(t))+\frac{w}{N}
\sum_j H(x_j(t-\tau)) + \xi_i(t)+I^{(e)}(t),
\hspace{1cm}\mbox{($i=1$ to $N$)}
\end{equation}
with
\begin{equation}
I^{(e)}(t)=A \Theta(t-t_{in}) \Theta(t_{in}+T_w-t).
\end{equation} 
Here $F(x)$ and $H(x)$ are functions of $x$, whose explicit
forms will be shown later 
[Eqs. (18), (19), (29), (30), (43) and (44)].
We have assumed uniform all-to-all couplings of $w$ 
and time delays of $\tau$.
The former assumption has been widely employed 
in many theoretical studies.
The latter assumption  may be justified 
in certain neural networks \cite{Salami03}. 
White noises of $\xi_i(t)$ are given by $<\xi_i(t)>=0$
and $<\xi_i(t) \xi_j(t')>= \beta^2 \:\delta_{ij}\delta(t-t')$
with the noise intensity of $\beta$ \cite{Note2}.
An applied input of $I^{(e)}(t)$ given by Eq. (2)
triggers oscillations in ensembles when model parameters are appropriate
as will be shown in Sec III, $\Theta(t)$ denoting the Heaviside function,
$A$ the magnitude, $t_{in}$ the input time, 
and $T_w$ the spike width.

In DMA \cite{Hasegawa03}, the global variable is given by
\begin{eqnarray}
X(t)=\frac{1}{N} \sum_i x_i(t), 
\end{eqnarray}
with which we define means and correlation functions given by
\begin{eqnarray}
\mu(t)&=&<X(t)>, \\
\gamma(t,t')&=& \frac{1}{N} \sum_i <\delta x_i(t)\:\delta x_i(t')>,\\
\rho(t,t')&=& <\delta X(t) \:\delta X(t')>,
\end{eqnarray}
using $\delta x_i(t)=x_i(t)-\mu(t)$ and $\delta X(t)=X(t)-\mu(t)$.

In deriving equations of motion of means and variances, we have
assumed that the noise intensity is weak and that
the state variables obey the Gaussian distributions around
their means, as in Refs. \cite{Hasegawa03,Hasegawa03b}. 
Numerical simulations 
have shown that for weak noises, 
the distribution of the state variable of an active rotator model
nearly obeys the Gaussian distribution,
although for strong noises, its distribution
deviates from the Gaussian \cite{Tanabe01}. 
Similar behavior
has been reported also in FN \cite{Tuckwell98}\cite{Tanabe01}
and HH neuron models
\cite{Tanabe99}\cite{Tanabe01a}.

After some manipulations,
we get DEs for $\mu(t)$, $\gamma(t,t)$ and $\rho(t,t')$
given by (for details see Appendix A),
\begin{eqnarray}
\frac{d \mu(t)}{dt}&=& g_0(t) + w u_0(t-\tau) + I^{(e)}(t), \\
\frac{d \gamma(t,t)}{dt}&=& 2 g_1(t) \gamma(t,t) 
+ 2 w u_1(t-\tau) \rho(t,t-\tau)  + \beta^2,\\
\frac{d \rho(t,t)}{d t}
&=& 2 g_1(t)\rho(t,t)+2 w u_1(t-\tau)\rho(t,t-\tau)+\frac{\beta^2}{N},\\
\frac{d \rho(t,t-m\tau)}{d t}
&=& [g_1(t)+g_1(t-m\tau)] \rho(t,t-m\tau)
+ w u_1(t-(m+1)\tau)\rho(t,t-(m+1)\tau) \nonumber \\ 
&+& w u_1(t-\tau) \rho(t-\tau, t-m\tau)
+ \frac{\beta^2}{N} \Delta(m\tau),
\;\;\;\;\;\mbox{(for $m \geq 1$)}
\end{eqnarray}
with
\begin{eqnarray}
g_0(t)&=& \sum_{n=0}^{\infty} \frac{F^{(2n)}(t)}{n !} 
\left( \frac{\gamma(t,t)}{2} \right)^n, \\
g_1(t)&=& \sum_{n=0}^{\infty} \frac{F^{(2n+1)}(t)}{n !} 
\left( \frac{\gamma(t,t)}{2} \right)^n, \\
u_0(t)&=& \sum_{n=0}^{\infty} \frac{H^{(2n)}(t)}{n !} 
\left( \frac{\gamma(t,t)}{2} \right)^n, \\
u_1(t)&=& \sum_{n=0}^{\infty} \frac{H^{(2n+1)}(t)}{n !} 
\left( \frac{\gamma(t,t)}{2} \right)^n,
\end{eqnarray}
where $\Delta(x)=1$ for $x=0$ and 0 otherwise.
Equations (7)-(10) show that an equation of motion of $\rho(t,t-\tau)$
includes new terms of $\rho(t-\tau,t-\tau)$ and $\rho(t,t-2 \tau)$,
which arise from the non-Markovian property of SDDE.
The recursive structure of DEs for $\rho(t,t)$ is schematically expressed 
in Fig. 1,
where arrows express the mutual dependence: 
$\rho(t,t)$ depends on $\rho(t,t-\tau)$, 
and $\rho(t,t-\tau)$ depends on $\rho(t,t-2\tau)$
and $\rho(t-\tau,t-\tau)$, and so on.
Then DMA transforms the original $N$-dimensional SDDEs given by
Eqs. (1) and (2) to the infinite-dimensional deterministic DEs
given by Eqs. (7)-(10).

In actual numerical calculations,
we will adopt the level-$m$ approximation (AMM$m$)
in which DEs are terminated at the finite level $m$:
\begin{equation}
\rho(t,t-(m+1)\tau) = \rho(t,t-m\tau).
\end{equation}
As will be shown later in model calculations with changing $m$,
the calculated result converges at a rather small $m$ 
[Figs. 2(a) and 9(b)].

We note that the noise contribution is $\beta^2$
in Eq. (8) and it is $\beta^2/N$ in Eq. (9).
It is easy to get
\begin{eqnarray}
\rho(t,t)&=& \frac{\gamma(t,t)}{N},
\hspace{2cm}\mbox{for $w/\beta^2 \rightarrow 0$} \\
&=& \gamma(t,t).
\hspace{2cm}\mbox{for $\beta^2/w \rightarrow 0$}
\end{eqnarray}
Equation (16) is consistent with the central-limit theorem.
We will show later that
with varying model parameters, the ratio of
$\rho(t,t)/\gamma(t,t)$ is varied, which leads to
a change in the synchronization of ensembles [Eq. (27)].

DSs have been performed for Eqs. (1) and (2) 
by using the fourth-order Runge-Kutta method with
a time step of 0.01.
Initial values of variables at $t \in (-\tau, 0]$ are 
$x_i(t)=x^*$ for $i=1$ to $N$,
where $x^*$ is the stationary solution for $\beta=0$. 
The trial number of DSs to be reported 
in the next Sec. III
is $N_r=100$ otherwise noticed.
AMM calculations have been performed for
Eqs. (7)-(10) with Eq. (15) by using also
the fourth-order Runge-Kutta method with
a time step of 0.01.
Initial values are
$\mu(t)=x^*$
and $\gamma(t,t)=0$
at $t \in [-\tau, 0]$, and 
$\rho(t,t')=0$ at $t \in [-\tau, 0]$
and $t' \in [-\tau, 0]$ ($t \geq t'$).
All calculated quantities are dimensionless. 

\section{Model calculations}

\subsection{Linear model}

We first consider the linear (L) model given by
\begin{eqnarray}
F(x)&=& -a x, 
\hspace{2cm}\mbox{($a \geq 0$)}\\
H(x)&=&x.
\end{eqnarray}
The stability of the stationary
solution of Eq. (1), (18) and (19) with $N=1$ and $I^{(e)}(t)=0$
was discussed in Refs. \cite{Kuchler92}-\cite{Frank03}, 
in particular, with the use of the moment method by
Mackey and Nechaeva \cite{Mackey95}.
When Eqs. (18) and (19) are adopted,
Eqs. (7)-(10) become
\begin{eqnarray}
\frac{d \mu(t)}{dt}&=& - a \mu(t)+w \mu(t-\tau)+ I^{(e)}(t), \\
\frac{d \gamma(t,t)}{dt}&=& -2 a \gamma(t,t) + 2 w \rho(t,t-\tau) + \beta^2,\\
\frac{d \rho(t,t)}{dt}&=& -2a \rho(t,t)+2 w \rho(t,t-\tau)+\frac{\beta^2}{N},\\
\frac{d \rho(t,t-m\tau)}{dt}&=& -2a \rho(t,t-m\tau)
+ w \rho(t,t-(m+1)\tau) 
\nonumber \\
&+& w \rho(t-\tau,t-m\tau)
+ \left( \frac{\beta^2}{2} \right) \Delta(m\tau),
\;\;\;\;\;\mbox{(for $m \geq 1$)}
\end{eqnarray}
because $g_0(t)=-a \mu(t)$, $g_1(t)=-a$, 
$u_0(t)=\mu(t)$ and $u_1(t)=1$ in Eqs. (11)-(14).

For $\beta=0$ and $I^{(e)}(t)=0$, 
Eq. (20) has the stationary solution of $\mu^*=0$.
Linearizing $\mu(t)$ around $\mu^*$, we get the condition
for the stationary solution given by
\begin{equation}
\tau < \tau_c=\frac{cos^{-1}(a/w)}{\sqrt{w^2-a^2}},
\end{equation}
which is just the same as the $N=1$ case \cite{Mackey95}.

\vspace{0.5cm}
\noindent
{\bf N=1 case}

First we discuss model calculations
for $N=1$, for which the exact solution
of its stationary state is available.
From Eqs. (21) and (22), we get $\rho(t,t')=\gamma(t,t')$
in the case of $N=1$.
Solid curves in Fig. 2(a) express $\gamma^*$, the stationary
value of $\gamma(t,t)$, when the level $m$
in AMM$m$ is varied for 
$a=1$, $w=0.5$ and $\beta=0.001$, whereas 
dashed curves denote the exact result given by
\cite{Kuchler92}
\begin{equation}
\gamma^{*}=\left( \frac{w \:sinh(\tau d)-d}
{2d[w \:cosh(\tau d)-a]}\right) \;\beta^2,
\hspace{2cm}\mbox{for $\mid w \mid < a$}
\end{equation}
where $d=\sqrt{a^2-w^2}$.
Equation (25) yields 
$10^6\:\gamma^* =$ 1.0, 0.724, 0.635, 0.581 and 0.577
for $\tau=$ 0, 1, 2, 5 and 10, respectively.
Figure 2(a) shows that for $\tau=0$, the result of AMM agrees with 
the exact one for $m \geq 0$.
In the case of $\tau=1$, the result of AMM is larger than the exact one for
$m=0$, but the former is smaller than the latter for $m \geq 1$. 
This is the case also for $\tau=2$.
In contrast, in cases of $\tau=5$ and 10, the results of AMM are in good
agreement with the exact ones for $m \geq 1$. 
It is surprising that the results of AMM converge at 
a small $m$ ($\sim 1$).
Solid and dashed curves in Fig. 2(b) show the $\tau$ dependence of $\gamma^*$
of AMM6 and the exact result, respectively 
(hereafter we show results in AMM6).
The result of AMM is in a fairly good agreement with the exact one
for $\tau > 4$.

Figure 2(b) shows the $\tau$ dependence of $\gamma^*$.
The result of DMA is in good agreement with the exact
one for $\tau > 3$.
It is interesting to make a comparison with results
calculated by the small-delay
approximation (SDA) initiated in Ref. \cite{Guil99},
some details of SDA being given in appendix B.
The dotted curve in Fig. 2(b) expresses $\gamma^*$ calculated in SDA
for $w=0.5$, $\beta=0.001$ and $N=1$.
Although the result of SDA agrees with the exact one at 
very small $\tau$ ($ \sim 0$), it shows a significant deviation
from the exact one at $\tau > 2$ where $\gamma^*$ becomes 
negative violating its positive definiteness.

Solid and dashed curves in Fig. 2(c) express
the $w$ dependence of $\gamma^{*}$ of AMM and exact ones, respectively,
for various $\tau$ values with $\beta=0.001$ and $N=1$.
We note that an agreement between AMM and exact results is good
except for $w > 0.6$ with $\tau = 1 \sim 2$ 
and for $w < -0.8$ with $\tau = 2 \sim 10$.
From the results shown in Figs. 2(a)-2(c), we may say
that AMM is a good approximation for 
a large $\tau \;(\geq 4)$ and a small $\beta$.

The response of the Langevin model ($N=1$) will be discussed
to an applied spike of $I^{(e)}(t)$ given by Eq. (2)
with $A\;(=0.5)$, $t_{in}\;(=100)$, 
and $T_w\;(=10)$.
Figures 3(a) and 3(b) show the time courses of $\mu(t)$ 
and $\gamma(t,t)\;(=\rho(t,t)$), 
respectively, with a set of parameters of
$a= 1$, $w=0.5$, $\beta=0.001$ and $N=1$, an input 
spike of $I^{(e)}(t)$ being shown at the bottom of Fig. 3(a). 
When an input spike is applied at $t=100$, state variables of $x_i(t)$
are randomized because independent noises have been added since
$t=0$.
Solid and dashed curves in Fig. 3(a), which
denote results of AMM and DSs, 
respectively, are practically identical.
The dotted curve expressing $\mu(t)$ of SDA is in fairly good agreement 
with that of DSs for $\tau=1$, but the former completely
disagrees with the latter for $\tau \geq 2$.
We should note in Eqs. (20)-(23) that $\mu(t)$ is decoupled 
from $\gamma(t,t)$ and $\rho(t,t)$, then
$\gamma(t,t)$ is independent of an external input $I^{(e)}(t)$.
Figure 3(b) shows that time courses of $\gamma(t,t)$ 
of AMM and DS are almost identical for $\tau=0$.
For $\tau=1$, result of AMM is underestimated
compared to that of DS as discussed before.
However, an agreement of the result of AMM with
that of DS becomes better for $\tau \geq 4$.
Results of $\gamma(t,t)$ of DS at large
$t$ ($> 100$) are in good agreement with the exact 
stationary solution of $\gamma^*$ shown in Fig. 2(c).

\vspace{0.5cm}
\noindent
{\bf N $>$ 1 case}

We will discuss dynamics, in particular, 
the synchronization, of ensembles for $N > 1$.
In order to monitor the synchronization, we consider the 
quantity given by \cite{Hasegawa03}
\begin{equation}
R(t)=\frac{1}{N^2} \sum_i <[x_i(t)-x_j(t)]^2>
=2[\gamma(t,t)-\rho(t,t)].
\end{equation}
When all neurons are in the completely synchronous states,
we get $x_i(t)=X(t)$ for all $i$ and then $R(t)=0$.
In contrast, in the asynchronous states, we get
$R(t)=2(1-1/N)\gamma(t,t) \equiv R_0(t)$
because $\rho(t,t)=\gamma(t,t)/N$ [Eq. (16)].
The {\it synchronization ratio} $S(t)$ is defined by \cite{Hasegawa03}
\begin{eqnarray}
S(t)&=& 1-\frac{R(t)}{R_0(t)} 
= \left( \frac{N\:\rho(t,t)/\gamma(t,t)-1}{N-1} \right),
\end{eqnarray}
which becomes 1 (0) for completely synchronous (asynchronous) states.
The synchrony $\sigma_s$ of the ensemble is defined by 
\begin{equation}
\sigma_s = \overline{S(t)}
=\left( \frac{1}{t_2-t_1} \right) \int_{t_1}^{t_2} \;dt \:S(t),
\end{equation}
where the overline denotes the temporal average
between times $t_1$ (=2000) and $t_2$ (=3000).

Figures 4(a) and 4(b) show the time course of $\mu(t)$ and $S(t)$,
respectively, for various $w$ with $\tau=10$, $\beta=0.001$ and $N=10$,
solid and dashed curves denoting results of AMM and DS, respectively.
For $w=0$, $\mu(t)$ behaves as a simple relaxation process with the
relaxation time of $\tau_r=1/a=1$, while $S(t)$ is vanishing.
When a small, positive coupling of $w=0.4$ is introduced,
$\mu(t)$ shows the stair-like structure because of the
positive delayed feedback.
The synchronization ratio $S(t)$ for $w=0.4$ 
shows a gradual development as increasing $t$,
but the magnitude of its averaged value of $\sigma_s$ 
is very small ($\sim 0.015$).
When $w$ is more increased to $w=0.8$, the effective relaxation time for
$\mu(t)$ to return to the initial zero value becomes larger
and $\sigma_s$ becomes also larger.
For $w=1.0$, the effective relaxation time
becomes infinity and $\mu(t)$ remains at the finite 
value of $\mu(t)=0.455$.
For $w > 1$, the divergence in $\mu(t)$ is triggered by an input spike
and $S(t)$ tends to a fully synchronized state of $\sigma_s =1$.
On the contrary, for a small negative coupling of $w=-0.4$,
$\mu(t)$ shows an ostensibly quasi-oscillating state because of
negative delayed feedback.
With increasing the magnitude of negative $w$, the term showing
this quasi-oscillation becomes longer.
For $w < -1.2$, $\mu(t)$ shows a divergent oscillation
and $S(t)$ tends to saturate at unity for $t > 250$.

The $w$ dependence of $\sigma_s$, which
is the temporal average of $S(t)$, is  
depicted in Fig. 5(a), where
solid and dashed curves show results of AMM and DS
with 1000 trials, respectively:
error bars showing the root-mean-square (RMS) value of
DS are within the radius of circles.
Although $\sigma_s$ is very small for $\mid w \mid < 0.9$,
it is suddenly increased as $\mid w \mid$ approaches the unity, 
where the divergence of the autonomous oscillation is induced
as shown in Fig. 4(a).

The delay time $\tau$ plays an important role, as discussed before 
in the case of $N=1$ [Fig. 2(b)].
Figure 5(b) shows the $\tau$ dependence of $\sigma_s$ calculated
by AMM (the solid curve) and DS
with 1000 trials (the dashed curve)
for $w=0.5$, $\beta=0.001$ and $N=10$.
We note that $\sigma_s=0.091$ for $\tau=0$
is rapidly decreased with increasing $\tau$ from zero,
while it is almost constant at $\tau > 4$.
This $\tau$ dependence of $\sigma_s$ resembles that of $\gamma^*$ for $N=1$
shown in Fig. 2(a).

We have so far fixed the size of $N$, which is now changed.
Figure 5(c) shows the $N$ dependence of $\sigma_s$ calculated
in AMM (the solid curve) and DS (the dashed curve)
for $\tau=10$, $w=1.0$ and $\beta=0.001$.
For $N=2$, the synchronization of $\sigma_s \sim 0.963$ is nearly 
complete.
With increasing $N$, however, $\sigma_s$ is gradually decreased:
$\sigma_s$= 0.824 and 0.340 for $N=10$ and 100, respectively.

Model calculations have shown that in linear Langevin ensembles
with appropriate model parameters,
an applied spike induces oscillations with divergent amplitudes.
This is contrast with the nonlinear Langevin ensembles
where stable oscillations with finite amplitudes
are possible, as will be shown in the following Sec. IIIB.

\subsection{Nonlinear model}

Next we consider the nonlinear (NL) model in which
$F(x)$ and $H(x)$ in Eq. (1) are given by
\begin{eqnarray}
F(x)&=& - a x, \\
H(x)&=&x - b x^3.
\hspace{2cm}\mbox{($a \geq 0, \;b > 0$)}
\end{eqnarray}
The NL model given by Eqs. (1), (29) and (30) 
with $a=0$, $b=1$, $I^{(e)}(t)=0$ and $N=1$ has been
discussed in Ref. \cite{Guil00}.
With the use of Eqs. (29) and (30), Eqs. (7)-(10) become
\begin{eqnarray}
\frac{d \mu(t)}{dt}&=& - a \mu(t)+w u_0(t-\tau)+ I^{(e)}(t), \\
\frac{d \gamma(t,t)}{dt}&=& - 2 a \gamma(t,t) 
+ 2 w u_1(t-\tau)\rho(t,t-\tau) + \beta^2,\\
\frac{d \rho(t,t)}{d t}
&=& -2 a \rho(t,t)+2 w u_1(t-\tau)\rho(t,t-\tau)+\frac{\beta^2}{N},\\
\frac{d \rho(t,t-m\tau)}{d t}
&=& -2 a \rho(t,t-m\tau)
+ w u_1(t-(m+1)\tau)\rho(t,t-(m+1)\tau) \nonumber \\
&+& w u_1(t-\tau) \rho(t-\tau,t-m\tau)
+\frac{\beta^2}{N} \Delta(m\tau),
\;\;\;\;\;\mbox{(for $m \geq 1$)}
\end{eqnarray}
with
\begin{eqnarray}
u_0(t)&=& \mu(t)-b \mu(t)^3- 3 b \mu(t)\gamma(t,t), \\
u_1(t)&=&1 - 3 b \mu(t)^2- 3 b \gamma(t,t).
\end{eqnarray}
For $\beta=0$ and $I^{(e)}(t)=0$, Eq. (31) has the stationary solution
given by
\begin{eqnarray}
\mu^* &=& 0,
\hspace{3cm}\mbox{for $w < a$} \\
&=& \pm \sqrt{\frac{w-a}{b w}}.
\hspace{2cm}\mbox{for $w > a$}
\end{eqnarray}
Linearizing Eq. (31) around $\mu^*$, we get the condition
for the stable stationary solution given by
\begin{eqnarray}
\tau &<& \tau_{c1}=\frac{cos^{-1}(a/w)}{\sqrt{(w^2-a^2)}},
\hspace{3cm}\mbox{for $w < a$} \\
&<& \tau_{c2}=\frac{cos^{-1}[a/(3a-2w)]}{\sqrt{[(3a-2w)^2-a^2]}}.
\hspace{2cm}\mbox{for $w > 2 a$}
\end{eqnarray}
Figure 6 shows the calculated $w$-$\tau$ phase diagram of the NL model,
showing the non-oscillating (NOSC) and oscillating states (OSC)
with $a=1$ and $b=1/6$, which are adopted for a later comparison
with the nonlinear model given by Eqs. (43) and (44) in Sec. IV. 
When an external spike given by Eq. (2) 
is applied to the NOSC state,
the state is once excited and returns to the 
stationary state after the transient period,
as will be discussed shortly [Figs. 8(a)-8(d)].
On the contrary, when a spike
is applied to the OSC state, 
it induces the autonomous oscillation.

Adopting parameters of $w$ and $\tau$ values shown by circles in Fig. 6,
we have made calculations for the NL model by AMM and DS, 
whose results of $\mu(t)$ and $S(t)$
are depicted in 
Figs. 7(a) and 7(b), respectively, 
with $\beta=0.001$ and $N=10$,
solid and dashed curves expressing results of
AMM and DS, respectively.
Hereafter we show results of AMM6 whose 
validity for the NL model
will be confirmed later [Fig. 9(b)].
%
Figure 7(a) shows that with increasing $\tau$, $\mu(t)$ shows
the complicated time dependence due to delayed feedbacks.
The time course of $\mu(t)$ for $N=10$ is the same as that for $N=1$
(results not shown). As was discussed in Sec IIIA, 
$\gamma(t,t)$ and $\rho(t,t)$ in the L model are
independent of an input signal $I^{(e)}(t)$ because they are decoupled
from $\mu(t)$ in Eq. (20)-(23).  It is not the case
in the NL model, where $\mu(t)$, $\gamma(t,t)$ and $\rho(t,t)$
are coupled each other in Eqs. (31)-(34), and $S(t)$ depends on $I^{(e)}(t)$.
Figure 7(b) shows that, for example, in the case of $\tau=0$,
$S(t) \sim 0.95$ at $t \siml 100$ is suddenly decreased to $S(t) \sim 0$
at $t=100$ by an applied spike, and then it is gradually increased to 
the stationary value of $S^* \sim 1.0$ at $t > 1000$.
This trend is realized in all the cases shown in Fig. 7(b).
We note that an agreement of $S(t)$ between AMM and DS
is good for $\tau=0$, 5 and 10, but not good for $\tau=1$ and 2,
just as in the case of the L model [Fig. 2(a)].

Figures 8(a)-8(d) show 
the time courses of $\mu(t)$ and $S(t)$
when the $w$ value is changed along the horizontal
dotted line in Fig. 6,
solid and dashed curves denoting results of AMM and DS, respectively.
From comparisons among Figs. 4(a), 4(b), and 8(a)-8(d), we note that 
for $\mid w \mid \:\leq 0.8$, time courses of $\mu(t)$ and $S(t)$
of the NL model are similar to those of the L model. 
The difference between the L and NL models is, 
however, clearly realized in cases of $\mid w \mid \geq 1.2$.
For $w=-1.2$, $\mu(t)$ in the NL model 
oscillates with the bounded magnitude [Fig. 8(c)]
while $\mu(t)$ in the L model
oscillates with divergent magnitude [Fig. 4(a)]
although the oscillating period is the same ($T = 22$) 
for L and NL models. In contrast,  
$S(t)$ in the NL model oscillates [Fig. 8(d)] 
while $S(t)$ in the L model saturates at the unity [Fig. 4(b)].
For $w=1.2$, $\mu(t)$ in the NL model
starting from the stationary state with $\mu^*=1.0$,
is slightly modified by
an input spike applied at $t=100$ with 
a small magnitude of $S^*=0.024$ for $S(t)$,
whereas $\mu(t)$ in the L model shows an unbounded oscillation
and $S(t)$ saturates at $S^*=1$.  
Figure 8(a) shows that 
as increasing $w$ above 1.2, $\mu(t)$ shows a quasi-oscillation 
triggered by inputs, by which $S(t)$ is increased
at $110 \siml t \siml 130$.
For $w > 2.0$, the autonomous 
oscillation with a period of $T = 22$ 
is induced and $S(t)$ is also oscillating 
with a period of $T = 11$. 

As discussed above, 
the oscillation is triggered by an input spike when parameters
are appropriate.
In order to study the transition between
the NOSC and OSC states in more detail, 
we have calculated the quantity $\sigma_o$ defined by\cite{Cartwright00}
\begin{eqnarray}
\sigma_o &=& \overline{ \frac{1}{N} \sum_i [<x_i(t)^2>-<x_i(t)>^2] }, \\
&=& \overline{\mu(t)^2}- \overline{\mu(t)}^2 + \overline{\gamma(t,t)},
\end{eqnarray}
which becomes finite in the OSC state but vanishes
(or is small) in the NOSC state,
the overline denoting the temporal average [Eq. (28)].

Figure 9(a) shows $\sigma_o$ and $\sigma_s$ calculated with
changing $w$ from 1.6 to 2.4 along the horizontal dotted line in Fig. 6
with $\tau=10$, $\beta=0.001$ and $N=10$, solid and dashed curves
denoting results of AMM6 and DS, respectively.
The oscillation is triggered 
by an input spike when $w$ exceeds the critical
value of $w_c \;(\sim $ 2.02).
The transition is of the second order since
$\sigma_o$ is continuously increased as $(w-w_c)$ is increased. 
We should note that the peak in $\sigma_s$ shows 
the {\it fluctuation-induced} enhancement at $w \sim w_c$,
which arises from an increase 
in the ratio of $\rho(t,t)/\gamma(t,t)$
although both $\gamma(t,t)$ and $\rho(t,t)$ are increased.
When $w$ exceeds about 2.3, the oscillation becomes irregular,
which is expected to be a precursor of the chaotic state.

For calculations of 
the NL model given by Eqs. (29) and (30),
we have adopted AMM6, whose
validity is examined in 
Fig. 9(b) showing $\sigma_s$ for $1.8 \leq w \leq 2.6$
when the level $m$ is changed in AMM$m$:
note that the result of $m=6$ in Fig. 9(b)
is nothing but the AMM result of $\sigma_s$ in Fig. 9(a).
The critical coupling for the NOSC-OSC transition 
calculated by AMM$m$ for $m=1-3$ is
too large compared to that by DS ($w_c=2.02$)
shown in Fig. 9(a).
For $m=4$, we get a reasonable value of $w_c$, but
the peak value of $\sigma_s$ at $w=w_c$ is too small.
With furthermore increasing the $m$ value,
the $w$-dependence of $\sigma_s$ becomes in better agreement
with that of DS. 
The optimum value of $m$ is expected to depend on
the model parameters, the required accuracy and 
the ability of computer facility. 
Making a compromise among these factors,
we have decided to adopt AMM6 in all our calculations. 
This choice of $m=6$
has been confirmed by results of AMM which are
in good agreement with those of DS [Fig 9(a)].

Figure 9(c) expresses 
the $w$ dependence of $\sigma_o$ and $\sigma_s$ 
near the NOSC-OSC transition for a negative $w$
when the $w$ value is changed from -1.4 to -0.6 
along the horizontal dotted line in Fig. 6.
The oscillation is triggered 
by an input spike when $w$ is below the critical
value of $w_c \;(\sim $ -1.04).
The fluctuation-induced enhancement is
again realized in $\sigma_s$ calculated by AMM and DS.

The $N$-dependence of $\sigma_s$ has been studied for
various $w$ ($\sim \:2$) with $\tau=10$ and $\beta=0.001$, whose results 
are depicted in Fig. 10(a).
It is noted that $\sigma_s$ with $w=2.02$ is larger than that
with $w=2.04$ for all $N$ values, just as shown in Fig. 9(a).
With increasing $N$, $\sigma_s$ is gradually decreased for
all $w$ values.
Although RMS values of $\sigma_s$ in the NOSC state
are small, they become considerable in the OSC states, 
which is due to oscillations in $S(t)$ 
but not due to noises.
In contrast, the $N$-dependence of $\sigma_o$ is
very small for the parameters investigated (results not shown).

Figure 10(b) shows the $\beta$ dependence of $\sigma_o$ and $\sigma_s$
for $w$ =2.0, 2.02 and 2.04 with $\tau=10$ and $N=10$.
For $w=2.02$, which is just the critical
coupling for $\beta=0.001$ [Fig. 9(a)], 
$\sigma_s$ is gradually decreased with increasing $\beta$.
For $w=2.00 \;(\siml w_c)$, $\sigma_s$ is little decreased
with an increase in $\beta$.
In contrast, for $w=2.04$, $\sigma_s$
is first increased with increasing $w$ 
and has a broad peak at $\beta \sim \beta_c$ where $\beta_c=0.04$ 
in AMM and 0.06 in DS.
This broad peak in $\sigma_s$
may suggest that the OSC state is suppressed by noises
although $\sigma_o$ calculated
in DS remains finite at $\beta > \beta_c$, showing no signs for the
NOSC-OSC transition.  
It may be possible that the emergence of the oscillation
is not well represented by $\sigma_o$ defined 
by Eqs. (41) and (42) in the case of a large $\beta$.
The discrepancy between results of
AMM and DS becomes significant 
with increasing $\beta$, which is due to 
a limitation of AMM based on the weak-noise assumption.

It has been shown that
when model parameters of $w$, $\tau$, $\beta$ and $N$ are appropriate,
the stable oscillations with finte magnitudes
are induced in NL Langevin ensembles.
The fluctuation-induced enhancement is realized in the synchrony
near the second-order transition between the NOSC and OSC states.

\section{Conclusions and Discussions}

We may adopt a nonlinear Langevin model given by Eq. (1) with
\begin{eqnarray}
F(x)&=& -a x, 
\hspace{2cm}\mbox{($a \geq 0$)}\\
H(x)&=&sin(x),
\end{eqnarray}
which is referred to as the NL' model.  Equations (43) and (44)
were previously employed by Ikeda and Matsumoto \cite{Ikeda80}
for a study on chaos in time-delayed systems.
By using Eqs. (43) and (44), and the relation given by
\begin{eqnarray}
H^{(2n)}(t)&=&(-1)^n \;sin(x), \\
H^{(2n+1)}(t)&=&(-1)^n \;cos(x),
\end{eqnarray}
we get DEs given by Eqs. (31)-(34) but with
\begin{eqnarray}
u_0(t)&=&sin(\mu(t)) \;exp \left( -\frac{\gamma(t,t)}{2} \right), \\
u_1(t)&=&cos(\mu(t)) \;exp \left( -\frac{\gamma(t,t)}{2} \right), 
\end{eqnarray}
where all contributions from $n=0$ to $\infty$
in Eqs. (11) and (14) are included.
It is noted that Eq. (30) with $b=1/6$ is an
approximate form of Eq. (44) for a small $x$.
Correspondingly, $u_0(t)$ and $u_1(t)$ given by
Eqs. (35) and (36) are approximate
expressions of those given by Eqs. (47) and (48), respectively, 
for small $\mu(t)$ and $\gamma(t,t)$.

Figure 11 shows the $w$ dependence of $\sigma_o$ and
$\sigma_s$ of the NL' model.
The NOSC-OSC transition occurs at $w_c = 2.29$
above which $\sigma_o$ is continuously increased and where $\sigma_s$
has a peak.
Although the induced oscillation is regular for $2.29 \leq w \siml 2.64$,
it becomes irregular for $w \simg 2.64$, which may
lead to the bifurcation and chaos \cite{Ikeda80}.
We note from Fig. 9(a) and 11 
that the $w$ dependence of the NL model given by
Eqs. (29) and (30) is similar to that of the NL' model 
given by Eqs. (43) and (44) 
although the critical coupling for the NOSC-OSC 
is different between the two models.
The $w-\tau$ phase diagram for NL' Langevin model is almost
the same as that for NL Langevin ensembles shown in Fig. 6.
Recently the phase diagram has been experimentally obtained
for a coupled {\it pair} of the plasmodium of the slime mold,
{\it physarum polycephalum}, where the coupling strength
and delay time are systematically controlled \cite{Takamatsu00}. 
The observed phase diagram is not dissimilar to
our $w-\tau$ phase diagrams for NL and NL' Langevin models
as far as unentrained and in-phase oscillating states are
concerned.

Quite recently, Huber and Tsimring (HT) \cite{Huber03} have discussed
an alternative nonlinear Langevin ensembles given by Eq. (1) with
\begin{eqnarray}
F(x)&=& x-x^3, \\
H(x)&=& x,
\end{eqnarray}
for $I^{(e)}(t)=0$, which expresses interconnected
bistable systems with delays \cite{Tsi01}. 
By using DSs and analytical methods based on the Gaussian and
dichotomous approximations, HT have discussed the 
coherence resonance and
multistability of the system.
When we apply our approximation to this nonlinear model, 
Eqs. (7)-(10) become 
\begin{eqnarray}
\frac{d \mu(t)}{dt}&=& \mu(t)-\mu(t)^3 - 3 \mu(t)\gamma(t,t)
+ w \mu(t-\tau)+I^{(e)}(t), \\
\frac{d \gamma(t,t)}{dt}&=& 2 [1-3 \mu(t)^2 -3\gamma(t,t)] \gamma(t,t) 
+ 2 w  \rho(t,t-\tau)  + \beta^2,\\
\frac{d \rho(t,t)}{d t}
&=& 2  [1-3 \mu(t)^2 -3\gamma(t,t)] \rho(t,t)
+2 w \rho(t,t-\tau)+\frac{\beta^2}{N},\\
\frac{d \rho(t,t-m\tau)}{d t}
&=& [g_1(t)+g_1(t-m\tau)] \rho(t,t-m\tau)
+ w \rho(t,t-(m+1)\tau) \nonumber \\ 
&+& \rho(t-\tau, t-m\tau)
+ \frac{\beta^2}{N} \Delta(m\tau),
\;\;\;\;\;\mbox{(for $m \geq 1$)}
\end{eqnarray}
where $g_1(t)=1-3 \mu(t)^2-3 \gamma(t,t)$.
In their Gaussian approximation, HT have employed
Eqs. (51) and (52) with $\rho(t, t-\tau)=0$,
discarding Eqs. (53) and (54).
It has been claimed that
the Gaussian approximation is not adequate
near the transition point between the ordered
and disordered states, although
a dichotomous theory yields a fairly good description.
This might be due to a neglect of
the higher-order contributions in Eqs. (52)-(54),
which are expected to play important roles, in particular,
near the transition point, as our calculations have shown 
[Fig. 9(b)].

In summary, we have proposed a semi-analytical
approach for a study of dynamics of stochastic ensembles 
described by linear and nonliner Langevin models with delays.
Advantages of our method are
(a) the synchronization in ensembles may be discussed by
taking into account correlations of local and global variables,
(b) the recursive DEs 
terminated at finite $m$ ($\sim 6$) yield fairly good results
compared to those of DSs, and
(c) our method is free from the magnitude of time
delays though the noise intensity is assumed to be weak,
which is complementary to SDA \cite{Guil99}.
The proposed method 
is expected to be useful not only to Langevin ensembles
but also to more general stochastic ensembles with delays. 
Although our method is applicable to the system 
with an arbitrary size of $N$, it is better applied 
to larger system because of its mean-field nature. 
It should be noted that the number of 
DEs to be solved for $N$-unit stochastic Langevin model
is $N N_r$ in DS with $N_r$ trials,
while it is $(m+3)$ in AMM$m$.
The ratio between the two numbers
becomes $N N_r/(m+3)\;\sim 1000$, for example, 
for $N=N_r=100$ and $m=6$. Actually this reflects on
the ratio of the speed for numerical computations
by using the two methods.
Taking these advantages of our method,
we have applied it to ensembles
described by FN neuron model with
delayed couplings to study their dynamics and synchronization,
which are reported in a following paper \cite{Hasegawa04b}.

\section*{Acknowledgments}
After a submission of the manuscript, the author learned
Ref. \cite{Huber03} from the referee, to whom he is much indebted.
This work is partly supported by
a Grant-in-Aid for Scientific Research from the Japanese 
Ministry of Education, Culture, Sports, Science and Technology.  


\appendix
\section{Derivation of Eqs. (7)-(10)}

Assuming that the noise intensity $\beta$ is small,
we express Eq. (1) in a Taylor expansion
of $\delta x_i$ as
\begin{eqnarray}
\frac{d x_i(t)}{dt}&=& \sum_{\ell =0}^{\infty}
\frac{F^{(\ell)}(t)}{\ell !} \delta x_i(t)^{\ell}
+ \frac{w}{N} \sum_j \sum_{\ell =0}^{\infty}
\frac{H^{(\ell)}(t-\tau)}{\ell !} \delta x_j(t-\tau)^{\ell} 
+ \xi_i(t) + I^{(e)}(t),
\end{eqnarray}
where $F^{(\ell)}(t)=F^{(\ell)}(\mu(t))$ and 
$H^{(\ell)}(t)=H^{(\ell)}(\mu(t))$.
Equations (3), (4) and (A1) yield
DE for means of $\mu(t)$ as
\begin{eqnarray}
\frac{d \mu(t)}{dt}&=& \frac{1}{N} \sum_i \sum_{\ell =0}^{\infty}
\frac{F^{(\ell)}(t)}{\ell !} <\delta x_i(t)^{\ell}> \nonumber \\
&+& \frac{w}{N^2} \sum_i \sum_j \sum_{\ell =0}^{\infty}
\frac{H^{(\ell)}(t-\tau)}{\ell !} <\delta x_j(t-\tau)^{\ell}>+ I^{(e)}(t).
\end{eqnarray}
When we adopt the Gaussian decoupling approximation,
averages higher than the second-order moments
in Eq. (A2) may be expressed in terms of the
second-order moments given by
\begin{eqnarray}
<\delta x_1,...,\delta x_{\ell}>
&=&\sum_{all \;parings} \Pi_{km} <\delta x_k \delta x_m>, 
\hspace{1cm}  
\mbox{for even $\ell$}, \nonumber \\
&=& 0,\hspace{5cm}\mbox{for odd $\ell$},
\end{eqnarray}
where the summation is performed for
all $(\ell-1)(\ell-3)....3\cdot1$ combinations.
With the use of the Gaussian decoupling approximation given by Eq. (A3), 
Eq. (A2) becomes
\begin{eqnarray}
\frac{d \mu(t)}{dt} &=& \frac{1}{N} \sum_i \sum_{n=0}^{\infty}
\frac{F^{(2n)}(t)}{(2n)!} B_{2n} <\delta x_i(t)^2>^n \nonumber \\
&+& \frac{w}{N^2} \sum_i \sum_j \sum_{n=0}^{\infty}
\frac{H^{(2n)}(t-\tau)}{(2n)!} B_{2n} <\delta x_j(t-\tau)^2>^n+ I^{(e)}(t),
\end{eqnarray}
where $B_{2n}=(2n-1)(2n-3)\cdot \cdot 3 \cdot 1$.
Adopting the mean-field approximation given by
\begin{equation}
<\delta x_i(t)^2>^n \:\simeq \:\gamma(t,t)^{n-1} <\delta x_i(t)^2>,
\end{equation}
we get
\begin{eqnarray}
\frac{d \mu(t)}{dt} &=& \sum_{n=0}^{\infty}
\frac{F^{(2n)}(t)}{n!} \left(\frac{\gamma(t,t)}{2} \right)^n
+ w \sum_{n=0}^{\infty}
\frac{H^{(2n)}(t-\tau)}{n!} \left(\frac{\gamma(t-\tau,t-\tau)}{2} \right)^n
+ I^{(e)}(t),
\end{eqnarray}
which yields Eq. (7), (11) and (13).

From Eqs. (A1) and (A4), we gt DEs for $d \delta x_i(t)/dt$ as
\begin{eqnarray}
\frac{d \delta x_i(t)}{dt} &=& \frac{d x_I(t)}{dt} - \frac{d \mu(t)}{dt} \\
&=& \sum_{n=0}^{\infty} F^{(2n+1)}(t) \frac{\delta x_i(t)^{2n+1}}{(2n+1)!}
+\sum_{n=0}^{\infty} F^{(2n)}(t) 
\left( \frac{\delta x_i(t)^{2n}}{(2n)!}
-\frac{\gamma(t,t)^n}{2^n \:n!} \right) \nonumber \\
&+& \frac{w}{N}\sum_j \sum_{n=0}^{\infty} H^{(2n+1)}(t-\tau) 
\frac{\delta x_j(t-\tau)^{2n+1}}{(2n+1)!} \nonumber \\
&+& \frac{w}{N} \sum_j \sum_{n=0}^{\infty} H^{(2n)}(t-\tau)
\left( \frac{\delta x_j(t-\tau)^{2n}}{(2n)!}
-\frac{\gamma(t-\tau,t-\tau)^n}{2^n\:n!}\right) + \xi_i(t).
\end{eqnarray}
By using Eqs. (5), (A3) and (A8), we get DEs for $d \gamma(t,t)/dt$ as
\begin{eqnarray}
\frac{d \gamma(t,t)}{d t} &=&
\frac{2}{N} \sum_i <\delta x_i(t) \frac{d \delta x_i(t)}{d t}> \\
&=& \frac{2}{N} \sum_i \sum_{n=0}^{\infty}
\frac{F^{2n+1}(t)}{(2n+1)!} <\delta x_i (t)^{2n+2}> \nonumber \\
&+& \frac{2w}{N^2} \sum_i \sum_j \sum_{n=0}^{\infty}
\frac{H^{2n+1}(t-\tau)}{(2n+1)!} 
<\delta x_i(t) \delta x_j (t-\tau)^{2n+1} >
+ \frac{2}{N} \sum_i <\delta x_i(t) \xi_i(t)> \nonumber \\
&=& \frac{2}{N} \sum_i \sum_{n=0}^{\infty}
\frac{F^{2n+1}(t)}{(2n+1)!} B_{2n+2} <\delta x_i(t)^2>^{n+1} \nonumber \\
&+& \frac{2w}{N^2} \sum_i \sum_j \sum_{n=0}^{\infty}
\frac{H^{2n+1}(t-\tau)}{(2n+1)!} B_{2n+2} <\delta x_i(t) \delta x_j(t-\tau) >
<\delta x_j(t-\tau)^2>^n \nonumber \\
&+& \beta^2.
\end{eqnarray}
With the use of the mean-field approximation given by Eq. (A5), Eq. (A10)
reduces to
\begin{eqnarray}
\frac{d \gamma(t,t)}{d t} 
&=& 2 \gamma(t,t) \sum_{n=0}^{\infty} 
\frac{F^{(2n+1)}(t)}{n!} \left(\frac{\gamma(t,t)}{2}\right)^n \nonumber\\
&+& 2 w \rho(t,t-\tau) \sum_{n=0}^{\infty} 
\frac{H^{(2n+1)}(t-\tau)}{n!} 
\left(\frac{\gamma(t-\tau,t-\tau)}{2}\right)^n + \beta^2,
\end{eqnarray}
leading to Eq. (8), (12) and (14).
Calculations of $d \rho(t,t)/dt$ and $d \rho(t,t-m\tau)/dt$
are similarly performed by using the relation:
\begin{eqnarray}
\frac{d \rho(t,t-m\tau)}{d t}
&=& \frac{1}{N^2} \sum_i \sum_j
<\delta x_i(t) \frac{d x_j(t-m\tau)}{dt}
+ \frac{d x_i(t)}{dt} x_j(t-m\tau) >.
\end{eqnarray}
In the process of calculating $d \rho(t,t-m\tau)/d t$, 
we get new correlation functions given by
\begin{eqnarray}
S(t,t-m\tau)&=&\frac{1}{N} \sum_i \:<\delta x_i(t) \xi_i(t-m\tau)>, \\
S(t-m\tau,t)&=&\frac{1}{N} \sum_i\:<\delta x_i(t-m\tau) \xi_i(t)>.
\end{eqnarray}
By using the method of steps 
in Ref. \cite{Frank03}, we get
\begin{equation}
S(t,t-m\tau)=S(t-m\tau,t)
= \left( \frac{\beta^2}{2} \right) \Delta(m\tau), 
\end{equation}
which leads to Eq. (10).

\section{The small-delay approximation}

We apply the small-delay approximation (SDA)
first proposed in Ref. \cite{Guil99} to our model
given by Eqs. (1) and (2).
When $\tau$ is small, we may expand Eq.(1) for $N=1$
as $x(t-\tau) \sim x(t) -\tau \:dx(t)/dt$ to get
\begin{equation}
\frac{d x(t)}{dt} \simeq F(x(t))
+ w \left( H(x(t))-\tau \:H'(x(t))\frac{d x(t)}{dt} \right)
+ \beta \eta(t)+I^{(e)}(t).
\end{equation}
Using Eq. (B1), we get DEs for $\mu(t)$ and $\gamma(t,t)$ given by
\begin{eqnarray}
\frac{d \mu(t)}{dt}&=& [1-w \tau h_1(t)]
[g_0(t)+w u_0(t)+I^{(e)}(t)], \\
\frac{d \gamma(t,t)}{dt}&=& 2[1-w \tau h_1(t)]
[g_1(t)+w u_1(t)] \gamma(t,t) + [1-w \tau h_1(t)]^2\beta^2,
\end{eqnarray}
where $h_1(t)=H^{'}(\mu(t))$.
For the L model given by Eqs. (1), (2), (18) and (19), 
Eqs. (B2) and (B3) become
\begin{eqnarray}
\frac{d \mu(t)}{dt}&=& (1-w \tau)[(-a +w) \mu(t)+I^{(e)}(t)], \\
\frac{\partial \gamma(t,t)}{\partial t}
&=& 2 (1 - w \tau)(-a + w) \gamma(t,t) + (1 - w \tau)^2 \beta^2.
\end{eqnarray}
The $\tau$ dependence of the stationary solution of $\gamma^*$
is shown by the dotted curve in Fig. 2(b).
The time course of $\mu(t)$ is plotted by doted curves in Fig. 3(a).

\begin{figure}
\caption{
The recursive structure of equations of motions
$\rho(t,t)$ in AMM, arrows denoting the mutual dependence
(see text).
}
\label{fig1}
\end{figure}

\begin{figure}
\caption{
(a) The stationary solution of $\gamma(t,t)$,
$\gamma^{*}$, of the linear (L) model
given by Eqs. (18) and (19) calculated in AMM$m$
with changing the level $m$ (solid curves)
and in the exact calculation (dashed lines)
for various $\tau$ with $a=1$, $w=0.5$, $\beta=0.001$ and $N=1$.
(b) The $\tau$ dependence of $\gamma^{*}$ 
in AMM6 (the solid curve), SDA (the dotted curve) 
and exact calculations (the dashed curve)
for $a=1$, $w=0.5$, $\beta=0.001$ and $N=1$.
 (c) The $w$ dependence of $\gamma^{*}$ 
for various $\tau$ 
in AMM6 (solid curves) and exact calculations (dashed curves)
for $\beta=0.001$ and $N=1$.
Results are multiplied by a factor of $10^6$, and those
of $\tau=5$, 2, 1, and 0 in (a) and (c) 
are successively shifted upwards by 1.
}
\label{fig2}
\end{figure}

\begin{figure}
\caption{
(color online).
The time course of (a) $\mu(t)$ and (b) $\gamma(t,t)$
of the L model 
for an applied single spike shown at the bottom 
frame in (a),
calculated in AMM (solid curves),
the small-delay approximation (SDA; dotted curves) and 
direct simulations (DS; dashed curves)
with $a=1$, $w=0.5$, $\beta=0.001$,
and $N=1$:
results of $\mu(t)$ for AMM and DS are
indistinguishable, and
results of SDA are shown only for $\tau \leq 2$
(see the text).
}
\label{fig3}
\end{figure}

\begin{figure}
\caption{
(color online).
Time courses of (a) $\mu(t)$ and (b) $S(t)$ 
of the L model
calculated by AMM (solid curves) 
and DS (dashed curves) for various $w$
with $\tau=10$, $\beta=0.001$ and $N=10$:
results of $\mu(t)$ for AMM and DS are
indistinguishable.
The chain curve at the bottom of (a) expresses an
applied input spike.
}
\label{fig4}
\end{figure}

\begin{figure}
\caption{
(a) The $w$ dependence of $\sigma_s$, 
the temporal average of $S(t)$, of the L model 
calculated in AMM (the solid curve) and
DS with $N_r=1000$ (the dashed curve)
for $\tau=10$, $\beta=0.001$ and $N=10$.
(b) The $\tau$ dependence of $\sigma_s$
for $w=0.5$, $\beta=0.001$ and $N=10$
calculated by AMM (the solid curve)
and DS with $N_r=1000$ (the dashed curve).
(c) The $N$ dependence of $\sigma_s$
for $w=0.5$, $\tau=10$ and $\beta=0.001$
calculated by AMM (the solid curve)
and DS (the dashed curve);
$N_r=100$ for $N=50$ and 100,
and $N_r=1000$ otherwise.
Error bars denote RMS values of DS.
}
\label{fig5}
\end{figure}

\begin{figure}
\caption{
The $w$-$\tau$ phase diagram of the nonlinear (NL) model
given by Eqs. (29) and (30) with $a=1$, $g=1/6$ and $\beta=0$,
showing the non-oscillating (NOSC) and
oscillating (OSC) states. 
Calculations whose results are depicted in Fig. 7, are performed
for sets of parameters shown by circles.
Along the horizontal dotted lines, the $w$ value is
changed for calculations shown 
in Figs. 8 and 9(a)-9(c) (see text).
}
\label{fig6}
\end{figure}

\begin{figure}
\caption{
(color online).
The time course of (a) $\mu(t)$ and (b) $S(t)$
of the NL model 
for an applied single spike shown at the bottom 
frame in (a),
calculated in AMM (solid curves) and 
DS (dashed curves)
for $a=1$, $w=1.0$, $\beta=0.001$,
and $N=10$:
results of $\mu(t)$ for AMM and DS are
indistinguishable.
}
\label{fig7}
\end{figure}

\begin{figure}
\caption{
(color online).
Time courses of (a) $\mu(t)$ and 
(b) $S(t)$  for $1.2 \leq w \leq 2.3$, and
(c) $\mu(t)$ and 
(d) $S(t)$ for $-1.2 \leq w \leq 1.0$,
of the NL model calculated by AMM (solid curves) 
and DS (dashed curves)
with $\tau=10$, $\beta=0.001$ and $N=10$:
results of $\mu(t)$ for AMM and DS are
indistinguishable.
Chain curves at bottoms of (a) and (c) express 
applied input spikes.
}
\label{fig8}
\end{figure}

\begin{figure}
\caption{
(a) The $w$ dependence of $\sigma_o$ and $\sigma_s$ 
for $1.6 \leq w \leq 2.4$ of the NL model, 
(b) The $w$ dependence of $\sigma_s$ 
for $1.8 \leq w \leq 2.6$ with different 
level $m$ in AMM$m$ (see text).
(c) The $w$ dependence of $\sigma_o$ and $\sigma_s$
for $-1.4 \leq w \leq -0.6$.
Solid and dashed curves in (a) and (c) 
express results of AMM6 and DS calculated
with $\tau=10$, $\beta=0.001$ and $N=10$.
Note that the result with $m=6$ in (b) corresponds to
the AMM result of $\sigma_s$ in (a).
Errors bars expressing RMS values are not
shown for a clarity of figures (see Fig. 10) 
}
\label{fig9}
\end{figure}

\begin{figure}
\caption{
(a) The $N$ dependence of $\sigma_s$ 
for $\tau=10$ and $\beta=0.001$, and
(b) the $\beta$ dependence of $\sigma_s$ 
for $\tau=10$ and $N=10$ of the NL model, calculated by
AMM (solid curves) and DS (dashed curves)
with $w=2.04$ (triangles), 2.02 (circles), 
2.0 (circles), 1.95 (diamonds) 
and 1.90 (inverted triangles).
Errors bars expressing RMS values of $\sigma_s$
of DS, are significant in the OSC state
because of the oscillation of $S(t)$.
}
\label{fig10}
\end{figure}

\begin{figure}
\caption{
The $w$ dependence of $\sigma_o$ and $\sigma_s$ 
of the NL' model given by Eqs. (43) and (44),
calculated by AMM (solid curves) 
and DS (dashed curves)
with $\tau=10$, $\beta=0.001$ and $N=10$ (see text).
Errors bars expressing RMS values are not
shown for a clarity of the figure.
}
\label{fig11}
\end{figure}

\end{document}